\documentclass[letterpaper]{article} 
\usepackage{aaai2027}  
\usepackage[hyphens]{url}  
\usepackage{graphicx} 
\usepackage{natbib}  
\usepackage{caption} 
\usepackage{booktabs}
\usepackage{amsmath}
\usepackage{amssymb}
\usepackage{algorithm}
\usepackage{algorithmic}

\usepackage{newfloat}
\usepackage{listings}
\DeclareCaptionStyle{ruled}{labelfont=normalfont,labelsep=colon,strut=off} 
\floatstyle{ruled}
\newfloat{listing}{tb}{lst}{}
\floatname{listing}{Listing}

\usepackage{booktabs}

\title{EEG-JEPA: Structured Latent Prediction for EEG Foundation Models}
\nocopyright
\author{
Jinhao Li\textsuperscript{\rm 1,2}\equalcontrib,
Zhiyuan Ma\textsuperscript{\rm 1,3}\equalcontrib,
Xueqiao Han\textsuperscript{\rm 1}\equalcontrib,
Zhongye Xia \textsuperscript{\rm 1,3},
Xinche Zhang \textsuperscript{\rm 1,3},
Shanghong Xie \textsuperscript{\rm 1,4},
Yixuan Liu \textsuperscript{\rm 1,3},
Yongjian Li \textsuperscript{\rm 1,3},
Runmin Gan \textsuperscript{\rm 1,3},
Tianlin Huo \textsuperscript{\rm 5}\corresponding,
Sen Song\textsuperscript{\rm 1,3}\corresponding
}

\affiliations{
\textsuperscript{\rm 1}Tsinghua Laboratory of Brain and Intelligence, Tsinghua University\\
\textsuperscript{\rm 2}School of Basic Medical Sciences, Tsinghua Medicine, Tsinghua University\\
\textsuperscript{\rm 3}
School of Biomedical Engineering, Tsinghua Medicine, Tsinghua University\\
\textsuperscript{\rm 4}Academy for Advanced Interdisciplinary Studies, Peking University\\
\textsuperscript{\rm 5}Department of Computer Science and Technology, Tsinghua University\\
huotianlin@gmail.com, songsen@tsinghua.edu.cn
}

\begin{document}

\maketitle

\begin{abstract}
Electroencephalography (EEG) foundation models aim to learn reusable representations from large-scale unlabeled recordings. A common pretraining strategy is masked waveform reconstruction, but applying supervision directly to noisy EEG may encourage models to recover predictable background activity, acquisition effects, and artifacts rather than neural structure that transfers across tasks. This raises a central question: what should an EEG foundation model predict to learn transferable representations? We introduce \textbf{EEG-JEPA}, a structured latent-prediction framework for EEG foundation modeling. Rather than reconstructing masked voltage samples, a masked context encoder and predictor infer contextual latent states produced by an exponential-moving-average target encoder that observes the complete input. EEG-JEPA organizes target design along three complementary dimensions: target content specifies what representation is predicted, target support specifies where prediction occurs over structured electrode--time regions through Neurotopology-Aware Multi-scale Electrode-Temporal Masking (N-MET), and target depth specifies at which encoder layers supervision is applied. Together, these designs shift EEG pretraining from recovering missing measurements to inferring latent states from structured electrode--time context. We evaluate EEG-JEPA through controlled objective comparisons, frozen multitask transfer, and full fine-tuning. Under the same backbone, pretraining corpus, and training duration, EEG-JEPA improves the 14-task frozen macro balanced accuracy from \textbf{40.49\% to 50.42\%} over CBraMod-style masked waveform reconstruction. Multi-source continuation further raises this result to \textbf{52.94\%}, the highest average among the EEG foundation models evaluated on EEG-FM-Bench. Under protocol-matched full fine-tuning, EEG-JEPA also improves the nine-task average balanced accuracy from \textbf{68.98\% to 70.65\%}. Code is available at \url{https://github.com/SWF-hao/EEG-JEPA-official}.
\end{abstract}


\section{Introduction}

Electroencephalography (EEG) provides a non-invasive window into brain activity and underpins applications in clinical monitoring, sleep analysis, brain--computer interfaces, and cognitive neuroscience \citep{obeid2016tuh,khalighi2016isruc,schalk2004bci2000,gifford2022things}. Large-scale self-supervised pretraining offers a promising route to learning representations that transfer across these diverse settings \citep{kostas2021bendr,jiang2024labram,wang2025cbramod}. Yet EEG presents distinctive challenges for foundation modeling. Scalp recordings have a low signal-to-noise ratio and non-stationary statistics, vary substantially across subjects, sites, and acquisition systems, and combine task-relevant neural dynamics with ongoing background activity, device effects, and physiological or environmental artifacts \citep{lotte2018review,melnik2017systems,jiang2019removal}. Moreover, downstream tasks depend on structure at different temporal and spatial scales: transient waveforms may be critical for event detection, rhythmic dynamics may characterize motor, cognitive, or sleep states, and informative activity may be distributed across multiple channels \citep{farwell1988talking,pfurtscheller1999event,khalighi2016isruc,DSAINet}. An effective EEG foundation model must therefore capture reusable neural structure across these scales rather than regularities specific to individual datasets or tasks. This requirement raises a central objective-design question: \emph{what should an EEG foundation model be trained to predict?}

Current EEG foundation models mainly rely on contrastive learning, masked signal reconstruction, discrete token prediction, representation alignment, or combinations of these objectives \citep{kostas2021bendr,yang2023biot,jiang2024labram,wang2024eegpt,wang2025cbramod,jiang2025neurolm}. Although these approaches have enabled large-scale EEG pretraining, they leave a fundamental question about what information the model is encouraged to retain. Reconstruction-based methods reward the recovery of all predictable signal components, while token-based methods inherit the information preserved by their tokenizers. For low-SNR EEG, both may emphasize background activity, subject- or device-specific patterns, and artifacts that are predictable but do not transfer reliably across tasks. Contrastive and alignment-based methods avoid exact signal recovery, but instead depend on the definitions of positive pairs, negative samples, and signal transformations \citep{kostas2021bendr,wang2024eegpt}. These assumptions are difficult to make task-independent: temporal shifts, channel perturbations, or frequency transformations may preserve the information required by one EEG task while altering that required by another. Hybrid objectives combine reconstruction and alignment but do not resolve which neural information should dominate the learned representation. Latent prediction offers an alternative by predicting contextual representations rather than raw observations, but existing EEG studies have not yet demonstrated consistently stronger transfer across diverse downstream tasks \citep{mohammadi2024eeg2rep}. This motivates a more fundamental question: how should latent targets be constructed to emphasize reusable neural structure across channels and temporal scales?

To answer this question, we introduce \textbf{EEG-JEPA}, a structured latent-prediction framework for EEG foundation modeling. Building on contextual latent prediction \citep{baevski2022data2vec,assran2023ijepa}, EEG-JEPA organizes latent-target design along three dimensions: target content specifies what representation is predicted, target support specifies where prediction occurs on the electrode--time lattice, and target depth specifies at which encoder layers supervision is applied. For target content, an exponential-moving-average encoder processes the complete EEG crop to produce contextual latent states, which a masked context encoder and predictor must infer from the visible electrode--time context. For target support, Neurotopology-Aware Multi-scale Electrode-Temporal Masking (N-MET) selects structured temporal, focal, regional, bilateral, and sensor-level regions rather than independent patches. For target depth, hierarchical prediction applies distinct targets at multiple encoder layers. Together, these designs shift EEG pretraining from recovering locally predictable voltage samples toward inferring structured latent states across the electrode--time lattice and representation hierarchy. Under the same backbone, Stage-1 corpus, and training duration, EEG-JEPA improves the 14-task frozen macro balanced accuracy from 40.49\% to 50.42\% over masked waveform reconstruction. Multi-source continuation further raises this result to 52.94\%, while protocol-matched full fine-tuning improves the nine-task average BA from 68.98\% to 70.65\%. Our contributions are threefold:
\begin{itemize}
    \item We formulate EEG foundation-model pretraining as a structured latent-prediction problem that jointly specifies contextual target content, neurotopology-aware electrode--time support, and representation depth, rather than reconstructing masked waveform samples.
    
    \item We instantiate this formulation with exponential-moving-average contextual targets, Neurotopology-Aware Multi-scale Electrode-Temporal Masking (N-MET), and depth-specific hierarchical prediction, requiring the model to infer latent states across temporal, focal, regional, bilateral, and sensor-level supports while directly supervising multiple encoder depths.
    
    \item We validate the proposed objective through strictly matched comparisons and progressive ablations, and demonstrate broad transferability through 14-task frozen transfer and nine-task full fine-tuning across heterogeneous EEG applications.
\end{itemize}

\begin{figure*}[t]
    \centering
    \includegraphics[width=\textwidth]{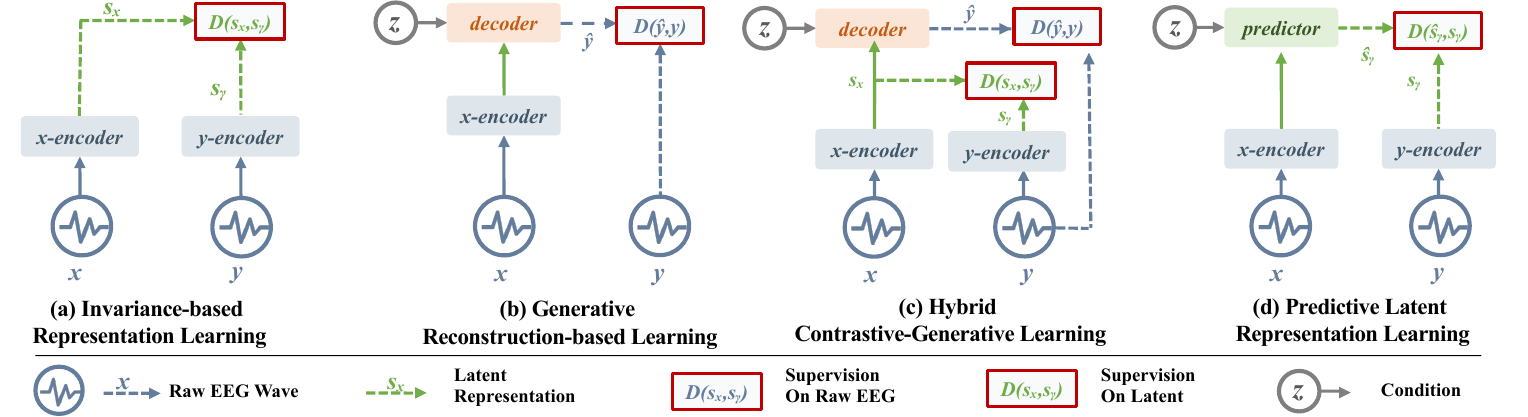}
    \caption{Comparison of major EEG self-supervised learning paradigms by supervision space. (a) Invariance-based methods align latent representations of two EEG views. (b) Generative methods reconstruct raw EEG signals from a corrupted input. (c) Hybrid methods combine waveform reconstruction with latent alignment. (d) Predictive latent methods infer the representation of a target EEG view from contextual representations and conditioning information, avoiding direct waveform reconstruction.}
    \label{fig:fm-framework}
\end{figure*}

\section{Related Work}

\paragraph{EEG Foundation Models.}
EEG foundation models seek transferable representations across tasks, datasets, and acquisition settings. BENDR uses contrastive predictive pretraining, while LaBraM, CBraMod, and EEGPT explore discrete neural-code prediction, masked waveform reconstruction, and masked modeling with momentum-target representation alignment, respectively \citep{kostas2021bendr,jiang2024labram,wang2025cbramod,wang2024eegpt}. Other work addresses heterogeneous inputs: BIOT supports cross-data biosignal learning, NeuroLM connects EEG with language for multitask learning, and REVE enables transfer across electrode montages \citep{yang2023biot,jiang2025neurolm,elouahidi2025reve}. Together, these studies establish the feasibility of EEG pretraining. However, because they differ in pretraining data, tokenization, channel handling, architecture, and learning objective, the effect of target design remains difficult to isolate. Existing work has not explicitly formulated EEG pretraining as joint target design over contextual content, electrode--time support, and representation depth.

\paragraph{Invariance and Reconstruction Objectives.}
Self-supervised objectives have mainly defined supervision through either agreement between related views or reconstruction of masked inputs. Contrastive and non-contrastive methods align representations of related views, using negative samples or architectural and statistical constraints to prevent representation collapse \citep{oord2018cpc,grill2020byol,bardes2022vicreg}. In time-series and EEG learning, these views are constructed through contextual consistency or temporal, spectral, and channel transformations \citep{eldele2021tstcc,yue2022ts2vec,banville2021uncovering,mohsenvand2020contrastive,kostas2021bendr}. However, transformations that preserve information for one EEG task may remove information required by another, while negative samples may not represent physiologically distinct states. Masked reconstruction instead predicts hidden samples, patches, or tokens from visible context, providing dense supervision at masked positions \citep{vincent2008dae,devlin2019bert,he2022mae,jiang2024labram,wang2025cbramod}. For EEG, however, such targets contain background rhythms, device effects, and artifacts alongside transferable neural structure. Temporal continuity and cross-channel redundancy may also allow masked content to be recovered from nearby observations without learning higher-level representations. Hybrid objectives combine representation alignment and reconstruction, but do not determine which variations should be ignored and which information should be retained. This limitation motivates predicting targets in a learned representation space.

\paragraph{Latent Predictive Learning.}
Latent predictive learning replaces agreement between predefined views and input reconstruction with the prediction of learned target representations from visible context. Data2vec uses an exponential-moving-average teacher that observes the complete input to generate contextual targets, whereas I-JEPA predicts representations of masked target regions from surrounding context and positional information \citep{baevski2022data2vec,assran2023ijepa}. EEG2Rep extends this approach to EEG by predicting masked latent representations rather than reconstructing raw signals \citep{mohammadi2024eeg2rep}. Together, these studies establish latent prediction as a viable alternative to direct signal reconstruction. However, they do not jointly define the target representation, the electrode--time positions to be predicted, and the encoder depths at which prediction targets are imposed. EEG-JEPA addresses this gap by explicitly specifying what is predicted, where prediction is performed, and at which encoder depths supervision is applied.

\section{EEG-JEPA: Structured Latent Predictive Learning}

\begin{figure}[t!]
    \centering
    \includegraphics[width=\linewidth]{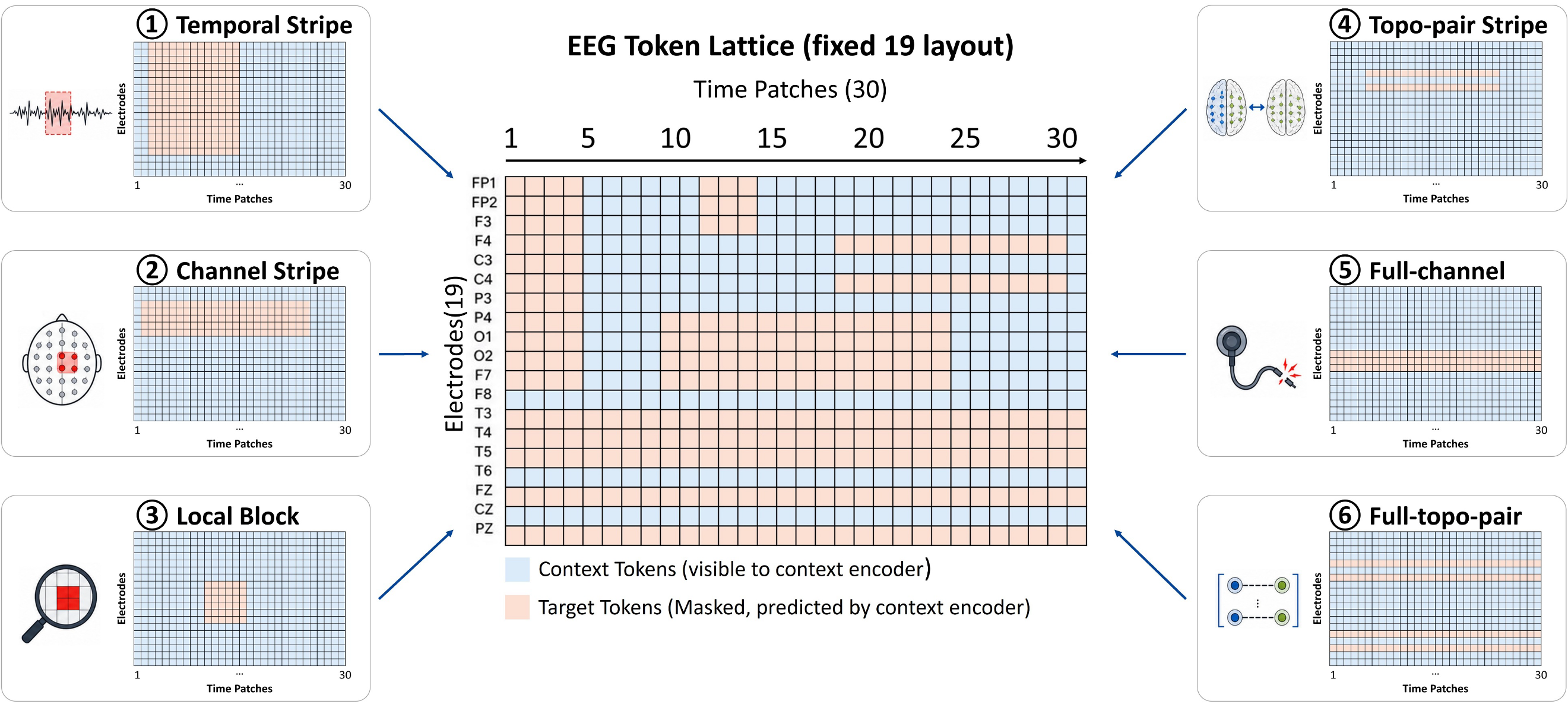}
    \caption{Neurotopology-Aware Multi-scale Electrode-Temporal Masking (N-MET) selects structured target locations on the electrode--time lattice.}
    \label{fig:nmet}
\end{figure}

\begin{figure*}[t]
    \centering
    \includegraphics[width=0.95\textwidth,height=0.34\textheight,keepaspectratio]{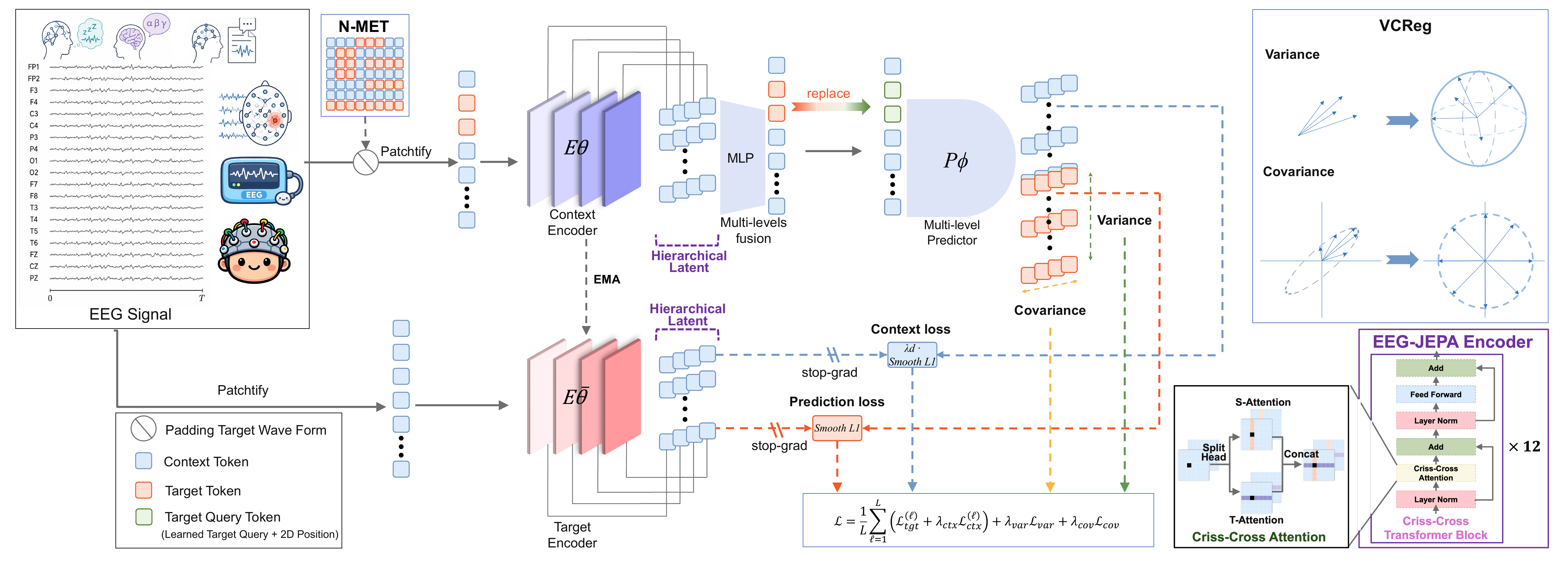}
    \caption{Overview of the EEG-JEPA pretraining framework. N-MET selects structured target locations on the electrode--time lattice. The masked crop is processed by the context encoder, while an EMA-updated target encoder receives the complete crop and produces stop-gradient contextual targets at multiple depths. A shared predictor combines hierarchical context features with learned position-conditioned target queries to predict the corresponding L3, L6, L9, and final-layer target representations. Training jointly minimizes latent prediction and context-consistency losses, together with variance and covariance regularization.}
    \label{fig:architecture}
\end{figure*}

\subsection{Framework Overview}

EEG-JEPA learns by predicting full-context latent states rather than masked
waveform values. Given a complete EEG crop $x$, N-MET first selects a set of
target positions $\mathcal{T}$ on the electrode--time lattice. The target
encoder observes the complete crop and produces reference representations at
these positions. In parallel, the context encoder processes the same crop
with the target waveform patches zeroed. The predictor receives only context
representations from visible positions, together with a positional query
identifying each target location, and estimates the corresponding
full-context representation. Gradients update the context encoder and
predictor, while the target encoder follows the context encoder through an
exponential moving average.

This prediction problem is governed by three coupled design decisions:
\textbf{target content} specifies which latent representation is predicted,
\textbf{target support} specifies which electrode--time positions must be
inferred, and \textbf{target depth} specifies where along the encoder hierarchy prediction is enforced. The subsections instantiate these decisions for EEG.

\subsection{Target Content: EMA Contextual States}

For a target electrode--time position $t$ and encoder depth $\ell$, EEG-JEPA
defines the prediction target as
\begin{equation}
    y_t^{(\ell)}
    =
    \mathrm{sg}\!\left(
    \operatorname{LN}\!\left(f_{\bar{\theta}}^{(\ell)}(x)_t\right)
    \right),
\end{equation}
where $f_{\bar{\theta}}$ is a target encoder that receives the complete crop,
$\operatorname{LN}$ controls target scale, and $\mathrm{sg}$ stops gradients
through the target branch. Unlike a waveform patch, $y_t^{(\ell)}$ is a
contextual representation: it describes position $t$ after information from
the complete crop has been integrated. The target parameters are updated from
the context encoder parameters by exponential moving average.

The context encoder $f_\theta$ instead receives
\begin{equation}
    \tilde{x}=\operatorname{ZeroMask}(x,\mathcal{T}),\qquad
    h_\theta^{(\ell)}=f_\theta^{(\ell)}(\tilde{x}),
\end{equation}
where $\operatorname{ZeroMask}$ replaces every target waveform patch with a
fixed all-zero vector before convolutional and spectral patch embedding.
Although the context encoder retains the complete electrode--time lattice,
only representations at visible positions $\mathcal{C}$ are gathered into
$z_{\mathcal C}$ and passed to the predictor. Representations at target
positions are therefore excluded from the prediction input.

Each target location is specified by
\begin{equation}
    m_t=e_q+p_t^{2\mathrm D},
\end{equation}
where $e_q$ is a learned target-query token and $p_t^{2\mathrm D}$ is a fixed
electrode--time sine--cosine positional embedding. The predictor estimates
\begin{equation}
    \hat{y}_{t}^{(\ell)}
    =g_\phi^{(\ell)}(z_{\mathcal C},m_t).
\end{equation}
The query reveals where a target is located but not its signal content.
Consequently, the prediction must be formed from visible EEG context, while
the full-input EMA encoder supplies the reference state.
\subsection{Target Support: Neurotopology-Aware Multi-scale
Electrode-Temporal Masking (N-MET)}

Having defined what is predicted, we next specify where prediction is
required. EEG patches form an electrode--time lattice whose axes have
different semantics. A target may represent a missing event interval, a
focal electrode--time pattern, a regional scalp field, a bilateral relation,
or an absent sensor. Independently sampled patches do not encode these
structures and, because of EEG's temporal continuity and cross-channel
redundancy, may be predictable from immediate neighbors alone.

N-MET constructs $\mathcal{T}$ by combining six electrode--temporal
primitives (Figure~\ref{fig:nmet} and Table~\ref{tab:nmet-primitives}. For
each crop, primitives are sampled until 40--55\% of valid tokens are selected;
padded positions are excluded from both attention and loss. The mixture
emphasizes temporal gaps while retaining focal, regional, bilateral, and
sensor-dropout supports, thereby creating prediction problems across multiple
spatial and temporal scales.

Topology-paired primitives probe bilateral correspondence and asymmetry, with full-pair masking requiring broader context than time-limited stripes. Full-channel masking simulates sensor loss. These masking patterns allow the model to learn temporal continuity, spatial correlation, and sensor robustness.

\begin{table}[t]
    \centering
    \scriptsize
    \setlength{\tabcolsep}{3pt}
    \begin{tabular}{p{0.2\linewidth}p{0.07\linewidth}p{0.66\linewidth}}
    \toprule
    Primitive & Mass & Design motivation \\
    \midrule
    Temporal stripe & 35.0\% & Missing event interval; infer morphology and rhythm continuity. \\
    Channel stripe & 17.5\% & Regional sensor loss; infer neighboring scalp fields. \\
    Local block & 17.5\% & Focal activity localized in electrode and time. \\
    Topo-pair stripe & 10.0\% & Bilateral correspondence and hemispheric asymmetry. \\
    Full channel & 15.0\% & Bad or absent electrode. \\
    Full topo-pair & 5.0\% & Bilateral dropout requiring global context. \\
    \bottomrule
    \end{tabular}
    \caption{N-MET primitives and conditional structure.}
    \label{tab:nmet-primitives}
\end{table}

\subsection{Target Depth: Hierarchical Supervision}

Target content and support define prediction at a particular encoder layer,
but supervision applied only to the final layer constrains only the endpoint
of the representation hierarchy. EEG-JEPA therefore predicts target states
at L3, L6, L9, and the final layer, spanning relatively local signal
structure and deeper contextual organization.

Concretely, the visible context representation $z_{\mathcal C}$ used above
is constructed by concatenating the 200-dimensional online states from these
four depths and fusing them with an
$800\!\rightarrow\!400\!\rightarrow\!200$ MLP. The fused states and target
queries are processed by a shared 100-dimensional predictor trunk. Four
independent $100\!\rightarrow\!200$ heads then estimate the corresponding
L3, L6, L9, and final-layer targets. The shared trunk couples information
across prediction depths, while the separate heads preserve depth-specific
target spaces.

Besides supervising the complete encoder, this design makes intermediate prefixes usable: frozen L3, L6, and L9 evaluations terminate the encoder at the corresponding block, providing a progressive accuracy--capacity trade-off.

\paragraph{Training objective.}
Let $\mathcal{D}=\{3,6,9,12\}$ denote the supervised depths and
$\mathcal{P}=\mathcal{T}\cup\mathcal{C}$ the valid target and context
positions. We define a unified latent prediction loss
\begin{equation}
    \mathcal{L}_{\mathrm{latent}}
    =
    \frac{1}{|\mathcal{D}|}
    \sum_{\ell\in\mathcal{D}}
    \underset{i\in\mathcal{P}}{\operatorname{mean}}\,
    w_i\,
    \rho\!\left(
        \hat{y}_i^{(\ell)}-y_i^{(\ell)}
    \right),
\end{equation}
where $\rho$ is the Smooth L1 loss and $w_i$ balances supervision over
masked targets and visible context positions. Invalid or padded positions are
excluded.

We additionally apply variance--covariance regularization (VCReg) to the
valid predictor outputs:
\begin{equation}
    \mathcal{L}_{\mathrm{VCReg}}
    =
    \lambda_{\mathrm{var}}\mathcal{L}_{\mathrm{var}}
    +
    \lambda_{\mathrm{cov}}\mathcal{L}_{\mathrm{cov}}.
\end{equation}
The variance component prevents collapsed feature dimensions, while the
covariance component reduces redundancy. The complete objective is
\begin{equation}
    \mathcal{L}
    =
    \mathcal{L}_{\mathrm{latent}}
    +
    \mathcal{L}_{\mathrm{VCReg}}.
\end{equation}
\section{Experiments}

\subsection{Pretraining Setup}
Stage 1 uses CBraMod's fixed 19-channel tokenization and
spatio-temporal encoder and trains all objective variants on
TUEG for 100 epochs \citep{wang2025cbramod}. All variants
share the same channel layout, patch definition, encoder
capacity, pretraining corpus, and training duration, enabling
controlled comparisons of raw and latent targets, independently
optimized and EMA target encoders, structured target support,
hierarchical prediction, context consistency, and VCReg.
Starting from the Stage-1 EEG-JEPA checkpoint, Stage 2
continues pretraining for 25 epochs on a mixture of TUEG,
TDBRAIN, and HBN (Table~\ref{tab:pretraining-data}). Each
training example is a valid 30-s crop. Stage 1 samples
1,115,752 crops per epoch, whereas Stage 2 samples 1,200,000
using weights of 75\%, 15\%, and 10\% for TUEG, TDBRAIN,
and HBN.

\begin{table}[!h]
\centering
\scriptsize
\setlength{\tabcolsep}{2.2pt}
\begin{tabular}{llrrr}
\toprule
Split & Sources & Ratio & Records & 30s crops \\
\midrule
Stage 1 & TUEG & 100\% & 39,758 & 6,535,247 \\
Stage 2 total & TUEG/TDBRAIN/HBN & 100\% & 44,486 & 6,648,828 \\
\quad TUEG part & TUEG & 75\% & 39,758 & 6,535,247 \\
\quad TDBRAIN part & TDBRAIN BVA & 15\% & 2,668 & 48,542 \\
\quad HBN part & HBN R8 & 10\% & 2,060 & 65,039 \\
\bottomrule
\end{tabular}
\caption{Two-stage trainnig corpus; ratios are stage-specific sampling weights.}
\label{tab:pretraining-data}
\end{table}


\subsection{Data Preprocessing}
Pretraining follows the CBraMod preprocessing protocol
\citep{wang2025cbramod}. Recordings are resampled to
200\,Hz, mapped to a canonical 19-channel montage with a
fixed channel order, segmented into 30-s crops, and divided
into non-overlapping 1-s patches. For downstream full
fine-tuning, each benchmark retains its prescribed channel set
and ordering rather than being mapped to the pretraining
montage, thereby preserving native recording configurations.
Further details on filtering, channel handling, quality control,
and segmentation are provided in the supplementary material.

\subsection{Evaluation Protocol}

We evaluate transfer under two regimes. For full fine-tuning, we follow the CBraMod protocol on nine datasets spanning clinical, motor-imagery, affective, sleep, stress, and imagined-speech tasks (Table~\ref{tab:finetuning-data}). Each dataset retains its prescribed channel set and ordering rather than being mapped to the canonical 19-channel pretraining montage. We compare EEG-JEPA with supervised EEG decoders and EEG foundation models \citep{lawhern2018eegnet,song2023eegconformer,wang2025cbramod}. External baseline results are taken from REVE \citep{elouahidi2025reve}.

For frozen transfer, we follow the 14-task EEG-FM-Bench multitask protocol, which covers clinical, sleep, motor-imagery, affective, workload, seizure, depression, and visual EEG tasks \citep{xiong2026eegfmbench}. We compare against the released foundation-model baselines spanning contrastive, reconstruction, and hybrid pretraining. We also include matched controls on the CBraMod backbone that vary the target-encoder update rule and target support. The encoder and its normalization statistics remain fixed, while mean-pooled token representations are passed to dataset-specific MLP heads that are trained jointly across tasks. We report balanced accuracy (BA) averaged over five downstream runs.

\begin{center}
\centering
\scriptsize
\setlength{\tabcolsep}{2.2pt}
\renewcommand{\arraystretch}{1.08}
\resizebox{\linewidth}{!}{%
\begin{tabular}{lllcc}
\toprule
Scale & Model & Pretraining paradigm & Params (M) & Macro BA (\%) \\
\midrule
Tiny & \textbf{EEG-JEPA-L3} & Latent predictive & 1.26 & \textbf{48.87} \\
\midrule
Small & \textbf{EEG-JEPA-L6} & Latent predictive & 2.47 & \textbf{49.06} \\
\midrule
Base & BIOT & Contrastive & 3.19 & 47.37 \\
 & \textbf{EEG-JEPA-L9} & Latent predictive & 3.68 & \textbf{49.70} \\
 & BENDR & Contrastive & 3.97 & 34.14 \\
\midrule
Large & \textbf{EEG-JEPA-final} & Latent predictive & 4.92 & \textbf{52.94} \\
 & CBraMod & Masked reconstruction & 4.92 & 40.49 \\
 & LaBraM & Masked token prediction & 5.82 & 44.91 \\
 & CSBrain & Masked reconstruction & 8.86 & 45.44 \\
\midrule
Huge & EEGPT & Masked + latent alignment & 25.29 & \underline{52.15} \\
 & REVE & Masked reconstruction & 69.19 & 51.50 \\
\bottomrule
\end{tabular}}
\captionof{table}{Frozen 14-task macro BA versus encoder size and pretraining paradigm.}
\label{tab:param-efficiency}
\end{center}

\begin{table}[t]
\centering

\resizebox{\columnwidth}{!}{%
\begin{tabular}{llccccc}
\toprule
\textbf{Task} & \textbf{Dataset} & \textbf{Ch.} &
\textbf{Duration} & \textbf{Samples} & \textbf{Orig. Hz}  &
\textbf{Classes} \\
\midrule
Abnormal detection  & TUAB     & 16 & 10s & 409,455 & 250 & 2 \\
Event type           & TUEV     & 16 & 5s  & 112,491 & 250 & 6 \\
Motor imagery        & PhysioMI & 64 & 4s  & 9,837   & 160 & 4 \\
Motor imagery        & BCI-2a   & 22 & 4s  & 5,184   & 250 & 4 \\
Emotion recognition  & FACED    & 32 & 10s & 10,332  & 250 & 9 \\
Sleep staging        & ISRUC    & 6  & 30s & 89,240  & 200 & 5 \\
Mental disorder      & Mumtaz   & 19 & 5s  & 7,143   & 256 & 2 \\
Mental stress        & MAT      & 20 & 5s  & 1,707   & 500 & 2 \\
Imagined speech      & BCI20-3  & 64 & 3s  & 6,000   & 256 & 5 \\
\bottomrule
\end{tabular}%

}
\caption{Full fine-tuning tasks and datasets.}
\label{tab:finetuning-data}
\end{table}

\subsection{Main Results}

\begin{table*}[!t]
\centering
\scriptsize
\setlength{\tabcolsep}{1.3pt}
\renewcommand{\arraystretch}{1.04}
\resizebox{\textwidth}{!}{%
\begin{tabular}{lcccccccccc}
\toprule
Method & TUAB & TUEV & PhysioMI & BCI-2a & FACED & ISRUC & Mumtaz & MAT & BCI20-3 & Avg. \\
\midrule
EEGNet & 76.42$\pm$0.36 & 38.76$\pm$1.43 & 58.14$\pm$1.25 & 44.82$\pm$0.94 & 40.90$\pm$1.22 & 71.54$\pm$1.21 & 92.32$\pm$1.04 & 67.70$\pm$1.16 & 44.13$\pm$0.96 & 59.41$\pm$0.37 \\
EEGConformer & 77.58$\pm$0.49 & 40.74$\pm$1.64 & 60.49$\pm$1.04 & 46.96$\pm$1.06 & 45.59$\pm$1.25 & 74.00$\pm$1.33 & 93.08$\pm$1.17 & 68.05$\pm$1.23 & 45.06$\pm$1.33 & 61.28$\pm$0.44 \\
SPaRCNet & 78.96$\pm$0.18 & 41.61$\pm$2.62 & 59.32$\pm$1.52 & 46.35$\pm$1.17 & 46.73$\pm$1.55 & 74.87$\pm$0.75 & 93.16$\pm$0.95 & 68.79$\pm$1.07 & 44.26$\pm$1.56 & 61.56$\pm$0.47 \\
ContraWR & 77.46$\pm$0.41 & 43.84$\pm$3.49 & 58.92$\pm$1.33 & 46.78$\pm$1.25 & 48.87$\pm$0.78 & 74.02$\pm$1.26 & 91.95$\pm$1.15 & 66.31$\pm$0.97 & 42.57$\pm$1.62 & 61.19$\pm$0.53 \\
CNN-Transformer & 77.77$\pm$0.22 & 40.87$\pm$1.61 & 60.53$\pm$1.18 & 46.00$\pm$1.08 & 46.97$\pm$1.32 & 73.63$\pm$0.87 & 93.05$\pm$0.68 & 67.79$\pm$2.68 & 45.33$\pm$0.92 & 61.33$\pm$0.45 \\
FFCL & 78.48$\pm$0.38 & 39.79$\pm$1.04 & 57.26$\pm$0.92 & 44.70$\pm$1.43 & 46.73$\pm$1.58 & 72.77$\pm$1.82 & 93.14$\pm$0.38 & 67.98$\pm$1.42 & 46.78$\pm$1.97 & 60.85$\pm$0.44 \\
ST-Transformer & 79.66$\pm$0.23 & 39.84$\pm$2.28 & 60.35$\pm$0.81 & 45.75$\pm$1.45 & 48.10$\pm$0.79 & 73.81$\pm$2.05 & 91.35$\pm$1.03 & 66.31$\pm$1.73 & 41.26$\pm$1.22 & 60.71$\pm$0.48 \\
\midrule
BIOT & 79.59$\pm$0.57 & 52.81$\pm$2.25 & 61.53$\pm$1.54 & 47.48$\pm$0.93 & 51.18$\pm$1.18 & 75.27$\pm$1.21 & 93.58$\pm$0.52 & 68.75$\pm$1.86 & 49.20$\pm$0.86 & 64.38$\pm$0.44 \\
LaBraM-Base & 81.40$\pm$0.19 & 64.09$\pm$0.65 & 61.73$\pm$1.22 & 48.69$\pm$0.85 & 52.73$\pm$1.07 & 76.33$\pm$1.02 & 94.09$\pm$0.79 & 69.09$\pm$1.25 & 50.60$\pm$1.55 & 66.53$\pm$0.31 \\
CBraMod & \underline{82.89$\pm$0.22} & \underline{66.71$\pm$1.07} & \underline{64.17$\pm$0.91} & \underline{51.38$\pm$0.66} & \underline{55.09$\pm$0.89} & \underline{78.65$\pm$1.10} & \underline{95.60$\pm$0.56} & \underline{72.56$\pm$1.32} & \underline{53.73$\pm$1.08} & \underline{68.98$\pm$0.31} \\
EEG-JEPA & \textbf{83.19$\pm$0.31} & \textbf{67.31$\pm$1.52} & \textbf{64.82$\pm$0.96} & \textbf{55.38$\pm$1.38} & \textbf{56.09$\pm$1.07} & \textbf{79.10$\pm$1.24} & \textbf{96.00$\pm$0.61} & \textbf{76.06$\pm$2.08} & \textbf{57.86$\pm$1.76} & \textbf{70.65$\pm$0.44} \\
\bottomrule
\end{tabular}}
\caption{Single-task full fine-tuning BA (\%, mean $\pm$ std).}
\label{tab:single-task-full-ft}
\end{table*}

\begin{figure}[!t]
    \centering
    \includegraphics[width=\linewidth]{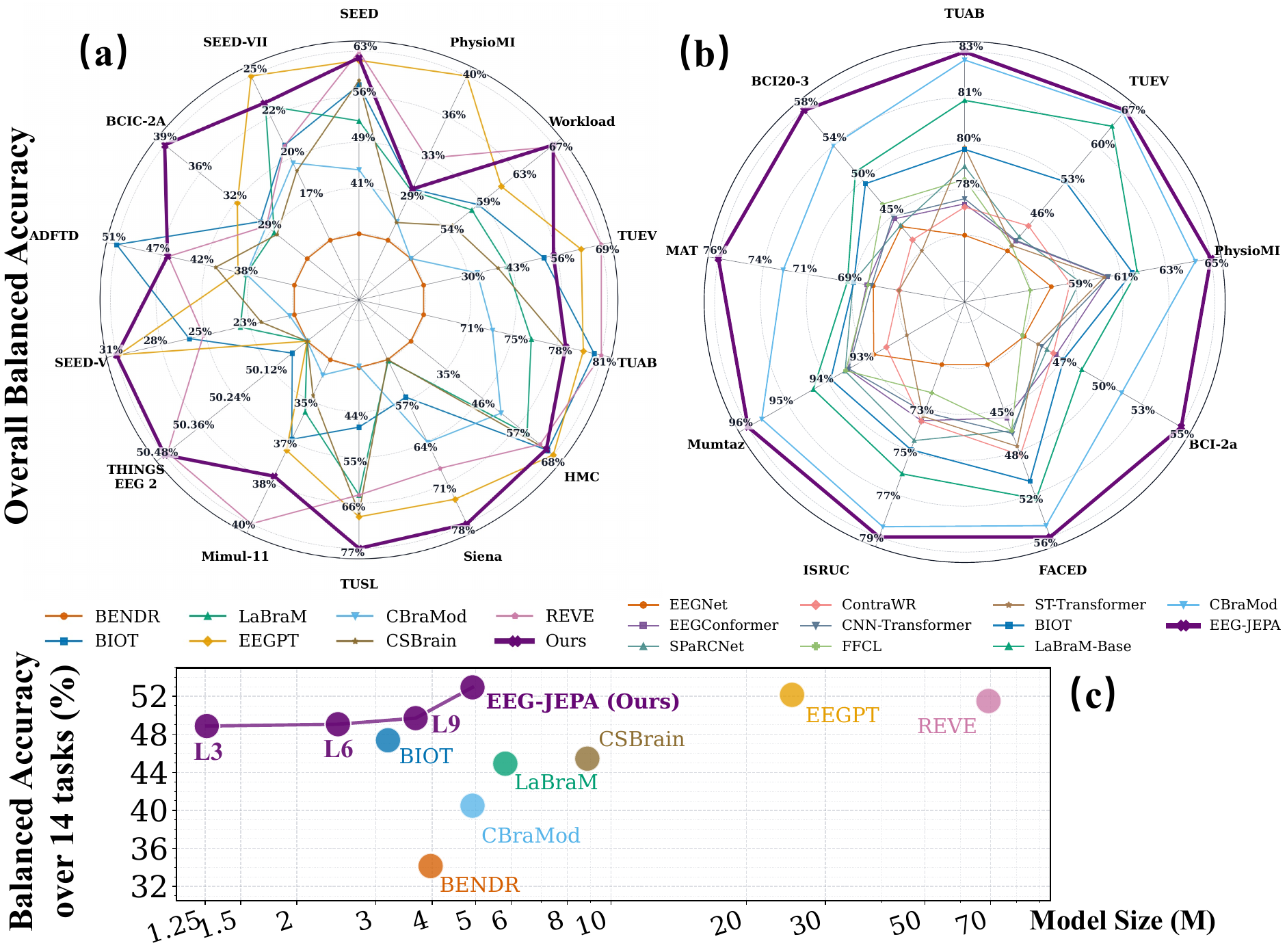}
    \caption{Results. (a) Frozen 14-task EEG foundation-model comparison on EEG-FM-Bench. (b) Full fine-tuning 9-task comparison. Different task suites are used because evaluations follow standard benchmarks; see Sec.~Evaluation Protocol for details. (c) Parameter-efficiency comparison for a.}
    \label{fig:foundation-radar}
\end{figure}

\paragraph{Frozen multitask transfer.}
Frozen-transfer results are summarized in Table~\ref{tab:param-efficiency} and Figure~\ref{fig:foundation-radar}c, while task-level BAs are shown in Figure~\ref{fig:foundation-radar}a, with detailed results provided in the appendix. With the encoder and its normalization statistics fixed, EEG-JEPA achieves the highest 14-task macro BA of 52.94\%, exceeding EEGPT and REVE by 0.79 and 1.44 percentage points, respectively, and ranks among the top two methods on 10 of the 14 tasks. Its performance across clinical, sleep, motor-imagery, affective, workload, seizure, depression, and visual EEG tasks shows that a single frozen representation transfers broadly without task-specific encoder adaptation.

\paragraph{Single-task full fine-tuning.}
Protocol-matched full fine-tuning gives EEG-JEPA the highest mean BA on all nine tasks, improving the nine-task average from 68.98\% for CBraMod to 70.65\%, an absolute gain of 1.67 percentage points (Table~\ref{tab:single-task-full-ft}). The largest improvements over CBraMod occur on BCI20-3 (+4.13 points), BCI-2a (+4.00), and MAT (+3.50), showing that the gains extend beyond clinical EEG to imagined speech, motor imagery, and stress.

\paragraph{Parameter-efficient transfer.}
Hierarchical supervision makes intermediate encoder prefixes usable at different model sizes. From the same pretrained encoder, the 1.26M-parameter L3, 2.47M-parameter L6, and 3.68M-parameter L9 prefixes achieve macro BAs of 48.87\%, 49.06\%, and 49.70\%, respectively, while the full 4.92M-parameter encoder reaches 52.94\%, compared with 40.49\% for the parameter-matched CBraMod encoder (Table~\ref{tab:param-efficiency}). A single pretrained model therefore provides an accuracy--size trade-off as additional encoder blocks are used.

\begin{center}
\scriptsize
\setlength{\tabcolsep}{2.0pt}
\resizebox{\linewidth}{!}{%
\begin{tabular}{lllcccccc}
\toprule
Objective variant & Space & Mask & Target & Multi-level & Context & VCReg & 2-stage & Macro BA \\
\midrule
CBraMod MAE control & Raw EEG & Random patch & -- &  &  &  &  & 40.49 $\pm$ 0.32 \\
Naive EEG-I-JEPA control & Latent & Random block & Ind. &  &  &  &  & 44.79 $\pm$ 0.91 \\
Naive EEG-V-JEPA control & Latent & Random block & EMA &  &  & \checkmark &  & 45.12 $\pm$ 0.43 \\
Naive EEG-V-JEPA control & Latent & N-MET & EMA &  &  & \checkmark &  & 46.47 $\pm$ 0.36 \\
+ multi-level prediction & Latent & N-MET & EMA & \checkmark &  & \checkmark &  & 50.13 $\pm$ 0.58 \\
+ context consistency & Latent & N-MET & EMA & \checkmark & \checkmark & \checkmark &  & 50.42 $\pm$ 0.58 \\
- VCReg control & Latent & N-MET & EMA & \checkmark & \checkmark &  &  & 50.09 $\pm$ 0.32 \\
Full EEG-JEPA & Latent & N-MET & EMA & \checkmark & \checkmark & \checkmark & \checkmark & \textbf{52.94 $\pm$ 0.30} \\
\bottomrule
\end{tabular}}
\captionof{table}{Progressive ablation of the EEG-JEPA objective design; BA is the 14-task macro average. Ind.\ denotes an independently optimized target encoder without EMA; EMA denotes an exponential-moving-average target encoder.}
\label{tab:controlled}
\end{center}

\paragraph{Ablation Study.}
Table~\ref{tab:controlled} evaluates structured latent prediction under a common Stage-1 setting with the same backbone, corpus, encoder capacity, and training duration. The CBraMod MAE control serves as the masked-waveform baseline, while independent- and EMA-target variants provide latent-prediction controls. We then replace random-block masking with N-MET to evaluate target support, add hierarchical prediction to evaluate target depth, and examine context consistency and VCReg as auxiliary objectives. Stage-2 continuation is reported separately because it changes the pretraining corpus.

The masked-waveform baseline achieves a 14-task macro BA of 40.49\%, while the independent- and EMA-target controls reach 44.79\% and 45.12\%. Under the matched EMA setting, N-MET improves performance from 45.12\% to 46.47\%, and hierarchical prediction provides the largest Stage-1 gain, raising it by 3.66 points to 50.13\%. Context consistency further increases performance to 50.42\%, while removing VCReg reduces it to 50.09\%. Multisource continuation then adds 2.52 points, reaching 52.94\%. These results show that EEG latent prediction benefits not only from learned targets, but also from structured target support and supervision across representation depths.

\subsection{Layerwise Representation Organization}

\begin{figure}[!t]
    \centering
    \includegraphics[width=\linewidth]{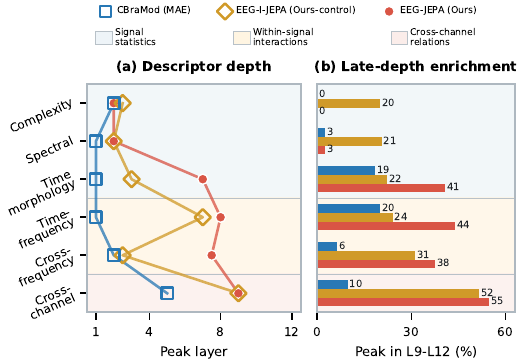}
    \caption{Matched descriptor organization across 12 frozen layers. (a) Median peak layer; curves connect families within each model and coincident markers share the same peak. (b) Fraction peaking in L9--L12. Bands denote signal statistics, within-signal interactions, and cross-channel relations.}
    \label{fig:semantic-hierarchy}
\end{figure}

We next examine whether the transfer gains of EEG-JEPA are accompanied by a different organization of information across encoder depth. Following the frozen representation audit of \citet{tang2026eegcapture}, we train ridge probes to predict 63 descriptors from six families on matched samples from five EEG-FM-Bench tasks. For each task--descriptor pair, the peak layer is selected after validation against shuffled and Gaussian controls, and comparisons use 158 pairs retained across the three models. We group the descriptor families into signal statistics, within-signal interactions, and cross-channel relations. Centered linear CKA complements the probe analysis by measuring changes in sample geometry across layers on matched inputs \citep{kornblith2019cka}.

\begin{figure}[!t]
    \centering
    \includegraphics[width=\linewidth]{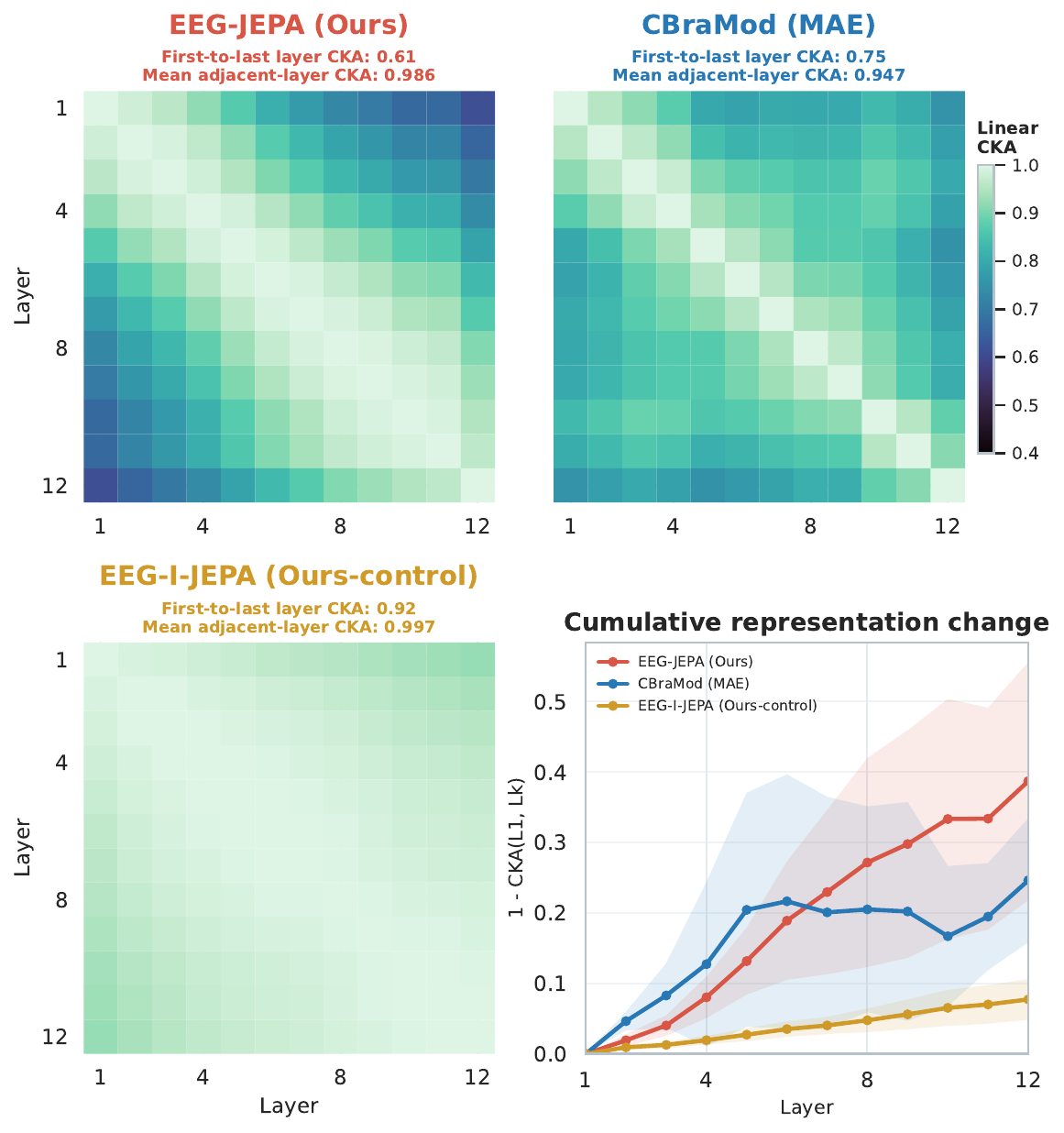}
    \caption{Task-averaged layerwise CKA and cumulative representation change on matched inputs.}
    \label{fig:representation-geometry}
\end{figure}

Distinct depth profiles emerge across the three models
(Figure~\ref{fig:semantic-hierarchy}a). CBraMod is front-loaded:
complexity, spectral, time-morphology, time-frequency, and
cross-frequency descriptors peak at L1--L2, whereas cross-channel
relations peak at L5. EEG-I-JEPA shifts time-frequency descriptors
to L7 and cross-channel relations to L9. EEG-JEPA shows a clearer
ordering, with complexity and spectral descriptors peaking at L2,
time morphology at L7, and within-signal interactions and
cross-channel relations around L8--L9.

Late-layer peak rates further show that EEG-JEPA does not uniformly
shift information deeper (Figure~\ref{fig:semantic-hierarchy}b).
Within L9--L12, the rates are 14.0\% for signal statistics, 41.5\%
for within-signal interactions, and 54.8\% for cross-channel
relations. Compared with EEG-I-JEPA, EEG-JEPA reduces late peaks for
signal statistics from 20.9\% to 14.0\%, while increasing them for
within-signal interactions from 26.8\% to 41.5\% and cross-channel
relations from 51.6\% to 54.8\%. CBraMod reaches only 14.6\% and
9.7\% for the latter two groups. This pattern indicates selective
late-depth enrichment of interaction and cross-channel information.

The CKA analysis clarifies how this depth organization develops
(Figure~\ref{fig:representation-geometry}). EEG-I-JEPA changes little
across depth, with a mean adjacent-layer CKA of 0.997 and an
L1--L12 CKA of 0.92. CBraMod exhibits larger changes
(0.947) but a smaller first-to-last change (0.75). EEG-JEPA combines
high adjacent-layer similarity (0.986) with the largest first-to-last
change (0.61), as its distance from L1 accumulates throughout the
encoder. Thus, small changes between neighboring layers build into
substantial reorganization across depth.

Together, the probe and CKA results show that EEG-JEPA develops a
selective depth organization: basic signal statistics remain most
accessible in shallow layers, whereas within-signal interactions and
cross-channel relations become more accessible in deeper
representations.

\section{Conclusion}
We introduced EEG-JEPA, a structured latent-prediction framework that jointly defines contextual target content, neurotopology-aware electrode--time support, and representation depth. N-MET structures prediction across multiple electrode--time supports, while hierarchical prediction supervises multiple encoder depths. Ablations identify target support and hierarchical prediction as the main sources of improvement, and layerwise analysis reveals increasing accessibility of interaction and cross-channel information at depth. EEG-JEPA achieves the highest frozen average among the foundation models evaluated on EEG-FM-Bench and improves full fine-tuning across nine tasks, demonstrating broad transfer across heterogeneous EEG applications.

\bibliography{references}

@article{lotte2018review,
  title={A review of classification algorithms for EEG-based brain--computer interfaces: a 10 year update},
  author={Lotte, Fabien and Bougrain, Laurent and Cichocki, Andrzej and Clerc, Maureen and Congedo, Marco and Rakotomamonjy, Alain and Yger, Florian},
  journal={Journal of neural engineering},
  volume={15},
  number={3},
  pages={031005},
  year={2018},
  publisher={iOP Publishing}
}

@inproceedings{jiang2025neurolm,
  title={NeuroLM: A universal multi-task foundation model for bridging the gap between language and EEG signals},
  author={Jiang, Wei-Bang and Wang, Yansen and Lu, Bao-Liang and Li, Dongsheng},
  booktitle={International conference on learning representations},
  volume={2025},
  pages={55436--55457},
  year={2025}
}

@inproceedings{mohammadi2024eeg2rep,
  title={Eeg2rep: enhancing self-supervised eeg representation through informative masked inputs},
  author={Mohammadi Foumani, Navid and Mackellar, Geoffrey and Ghane, Soheila and Irtza, Saad and Nguyen, Nam and Salehi, Mahsa},
  booktitle={Proceedings of the 30th ACM SIGKDD Conference on Knowledge Discovery and Data Mining},
  pages={5544--5555},
  year={2024}
}

@article{melnik2017systems,
  title={Systems, subjects, sessions: to what extent do these factors influence EEG data?},
  author={Melnik, Andrew and Legkov, Petr and Izdebski, Krzysztof and K{\"a}rcher, Silke M and Hairston, W David and Ferris, Daniel P and K{\"o}nig, Peter},
  journal={Frontiers in human neuroscience},
  volume={11},
  pages={150},
  year={2017},
  publisher={Frontiers Media SA}
}

@article{jiang2019removal,
  title={Removal of artifacts from EEG signals: a review},
  author={Jiang, Xiao and Bian, Gui-Bin and Tian, Zean},
  journal={Sensors},
  volume={19},
  number={5},
  pages={987},
  year={2019},
  publisher={MDPI}
}

@article{pfurtscheller1999event,
  title={Event-related EEG/MEG synchronization and desynchronization: basic principles},
  author={Pfurtscheller, Gert and Da Silva, FH Lopes},
  journal={Clinical neurophysiology},
  volume={110},
  number={11},
  pages={1842--1857},
  year={1999},
  publisher={Elsevier}
}

@article{farwell1988talking,
  title={Talking off the top of your head: toward a mental prosthesis utilizing event-related brain potentials},
  author={Farwell, Lawrence Ashley and Donchin, Emanuel},
  journal={Electroencephalography and clinical Neurophysiology},
  volume={70},
  number={6},
  pages={510--523},
  year={1988},
  publisher={Elsevier}
}

@article{DSAINet,
  title={Dsainet: An efficient dual-scale attentive interaction network for general eeg decoding},
  author={Ma, Zhiyuan and Li, Zeyuan and Qiu, Zihao and Li, Jinhao and Meng, Lingqin and Zhang, Xinche and Liu, Yixuan and Shen, Xinke and Song, Sen},
  journal={arXiv preprint arXiv:2604.18095},
  year={2026}
}

@article{oord2018cpc,
  title={Representation learning with contrastive predictive coding},
  author={Oord, Aaron van den and Li, Yazhe and Vinyals, Oriol},
  journal={arXiv preprint arXiv:1807.03748},
  year={2018}
}

@inproceedings{grill2020byol,
  title = {Bootstrap Your Own Latent: A New Approach to Self-Supervised Learning},
  author = {Grill, Jean-Bastien and Strub, Florian and Altch{\'e}, Florent and Tallec, Corentin and Richemond, Pierre and Buchatskaya, Elena and Doersch, Carl and Pires, Bernardo Avila and Guo, Zhaohan and Azar, Mohammad Gheshlaghi and Piot, Bilal and Kavukcuoglu, Koray and Munos, R{\'e}mi and Valko, Michal},
  booktitle = {Advances in Neural Information Processing Systems},
  year = {2020}
}

@inproceedings{bardes2022vicreg,
  title = {{VICReg}: Variance-Invariance-Covariance Regularization for Self-Supervised Learning},
  author = {Bardes, Adrien and Ponce, Jean and LeCun, Yann},
  booktitle = {International Conference on Learning Representations},
  year = {2022}
}

@inproceedings{vincent2008dae,
  title = {Extracting and Composing Robust Features with Denoising Autoencoders},
  author = {Vincent, Pascal and Larochelle, Hugo and Bengio, Yoshua and Manzagol, Pierre-Antoine},
  booktitle = {Proceedings of the International Conference on Machine Learning},
  year = {2008}
}

@inproceedings{devlin2019bert,
  title = {{BERT}: Pre-training of Deep Bidirectional Transformers for Language Understanding},
  author = {Devlin, Jacob and Chang, Ming-Wei and Lee, Kenton and Toutanova, Kristina},
  booktitle = {Proceedings of the North American Chapter of the Association for Computational Linguistics},
  year = {2019}
}

@inproceedings{he2022mae,
  title = {Masked Autoencoders Are Scalable Vision Learners},
  author = {He, Kaiming and Chen, Xinlei and Xie, Saining and Li, Yanghao and Doll{\'a}r, Piotr and Girshick, Ross},
  booktitle = {Proceedings of the IEEE/CVF Conference on Computer Vision and Pattern Recognition},
  year = {2022}
}

@inproceedings{assran2023ijepa,
  title = {Self-Supervised Learning from Images with a Joint-Embedding Predictive Architecture},
  author = {Assran, Mahmoud and Duval, Quentin and Misra, Ishan and Bojanowski, Piotr and Vincent, Pascal and Rabbat, Michael and LeCun, Yann and Ballas, Nicolas},
  booktitle = {Proceedings of the IEEE/CVF Conference on Computer Vision and Pattern Recognition},
  year = {2023}
}

@inproceedings{eldele2021tstcc,
  title = {Time-Series Representation Learning via Temporal and Contextual Contrasting},
  author = {Eldele, Emadeldeen and Ragab, Mohamed and Chen, Zhenghua and Wu, Min and Kwoh, Chee-Keong and Li, Xiaoli and Guan, Cuntai},
  booktitle = {Proceedings of the International Joint Conference on Artificial Intelligence},
  year = {2021}
}

@inproceedings{yue2022ts2vec,
  title = {{TS2Vec}: Towards Universal Representation of Time Series},
  author = {Yue, Zhihan and Wang, Yujing and Duan, Juanyong and Yang, Tian and Huang, Congrui and Tong, Yunhai and Xu, Bixiong},
  booktitle = {Proceedings of the AAAI Conference on Artificial Intelligence},
  year = {2022}
}

@article{banville2021uncovering,
  title = {Uncovering the Structure of Clinical {EEG} Signals with Self-Supervised Learning},
  author = {Banville, Hubert and Chehab, Omar and Hyv{\"a}rinen, Aapo and Engemann, Denis A. and Gramfort, Alexandre},
  journal = {Journal of Neural Engineering},
  volume = {18},
  number = {4},
  pages = {046020},
  year = {2021}
}

@inproceedings{mohsenvand2020contrastive,
  title={Contrastive representation learning for electroencephalogram classification},
  author={Mohsenvand, Mostafa Neo and Izadi, Mohammad Rasool and Maes, Pattie},
  booktitle={Machine learning for health},
  pages={238--253},
  year={2020},
  organization={PMLR}
}

@article{kostas2021bendr,
  title = {{BENDR}: Using Transformers and a Contrastive Self-Supervised Learning Task to Learn from Massive Amounts of {EEG} Data},
  author = {Kostas, Demetres and Aroca-Ouellette, St{\'e}phane and Rudzicz, Frank},
  journal = {Frontiers in Human Neuroscience},
  volume = {15},
  pages = {653659},
  year = {2021}
}

@inproceedings{yang2023biot,
  title = {{BIOT}: Biosignal Transformer for Cross-Data Learning in the Wild},
  author = {Yang, Chaoqi and Westover, M. Brandon and Sun, Jimeng},
  booktitle = {Advances in Neural Information Processing Systems},
  year = {2023}
}

@inproceedings{jiang2024labram,
  title = {Large Brain Model for Learning Generic Representations with Tremendous {EEG} Data in {BCI}},
  author = {Jiang, Wei-Bang and Zhao, Li-Ming and Lu, Bao-Liang},
  booktitle = {International Conference on Learning Representations},
  year = {2024}
}

@article{wang2024eegpt,
  title={Eegpt: Pretrained transformer for universal and reliable representation of eeg signals},
  author={Wang, Guagnyu and Liu, Wenchao and He, Yuhong and Xu, Cong and Ma, Lin and Li, Haifeng},
  journal={Advances in Neural Information Processing Systems},
  volume={37},
  pages={39249--39280},
  year={2024}
}

@inproceedings{wang2025cbramod,
  title = {{CBraMod}: A Criss-Cross Brain Foundation Model for {EEG} Decoding},
  author = {Wang, Jiquan and Zhao, Sha and Luo, Zhiling and Zhou, Yangxuan and Jiang, Haiteng and Li, Shijian and Li, Tao and Pan, Gang},
  booktitle = {International Conference on Learning Representations},
  year = {2025}
}

@inproceedings{elouahidi2025reve,
  title = {{REVE}: A Foundation Model for {EEG}---Adapting to Any Setup with Large-Scale Pretraining on 25,000 Subjects},
  author = {El Ouahidi, Yassine and Lys, Jonathan and Th{\"o}lke, Philipp and Farrugia, Nicolas and Pasdeloup, Bastien and Gripon, Vincent and Jerbi, Karim and Lioi, Giulia},
  booktitle = {Advances in Neural Information Processing Systems},
  volume = {38},
  year = {2025}
}

@article{xiong2026eegfmbench,
  title={Eeg-fm-bench: A comprehensive benchmark for the systematic evaluation of eeg foundation models},
  author={Xiong, Wei and Li, Jiangtong and Li, Jie and Zhu, Kun},
  journal={arXiv preprint arXiv:2508.17742},
  year={2025}
}

@article{schalk2004bci2000,
  title = {{BCI2000}: A General-Purpose Brain-Computer Interface ({BCI}) System},
  author = {Schalk, Gerwin and McFarland, Dennis J. and Hinterberger, Thilo and Birbaumer, Niels and Wolpaw, Jonathan R.},
  journal = {IEEE Transactions on Biomedical Engineering},
  volume = {51},
  number = {6},
  pages = {1034--1043},
  year = {2004}
}

@article{obeid2016tuh,
  title = {The Temple University Hospital {EEG} Data Corpus},
  author = {Obeid, Iyad and Picone, Joseph},
  journal = {Frontiers in Neuroscience},
  volume = {10},
  pages = {196},
  year = {2016}
}

@article{khalighi2016isruc,
  title = {{ISRUC-Sleep}: A Comprehensive Public Dataset for Sleep Researchers},
  author = {Khalighi, Sirvan and Sousa, Teresa and Santos, Jos{\'e} Moutinho and Nunes, Urbano},
  journal = {Computer Methods and Programs in Biomedicine},
  volume = {124},
  pages = {180--192},
  year = {2016}
}

@article{gifford2022things,
  title = {Human {EEG} Recordings for 1,854 Concepts Presented in Rapid Serial Visual Presentation Streams},
  author = {Grootswagers, Tijl and Zhou, Ivy and Robinson, Amanda K. and Hebart, Martin N. and Carlson, Thomas A.},
  journal = {Scientific Data},
  volume = {9},
  pages = {3},
  year = {2022}
}

@article{lawhern2018eegnet,
  title = {{EEGNet}: A Compact Convolutional Neural Network for {EEG}-Based Brain-Computer Interfaces},
  author = {Lawhern, Vernon J. and Solon, Amelia J. and Waytowich, Nicholas R. and Gordon, Stephen M. and Hung, Chou P. and Lance, Brent J.},
  journal = {Journal of Neural Engineering},
  volume = {15},
  number = {5},
  pages = {056013},
  year = {2018}
}

@article{song2023eegconformer,
  title = {{EEG} Conformer: Convolutional Transformer for {EEG} Decoding and Visualization},
  author = {Song, Yonghao and Zheng, Qingqing and Liu, Bingchuan and Gao, Xiaorong},
  journal = {IEEE Transactions on Neural Systems and Rehabilitation Engineering},
  volume = {31},
  pages = {710--719},
  year = {2023}
}

@inproceedings{baevski2022data2vec,
  title = {data2vec: A General Framework for Self-Supervised Learning in Speech, Vision and Language},
  author = {Baevski, Alexei and Hsu, Wei-Ning and Xu, Qiantong and Babu, Arun and Gu, Jiatao and Auli, Michael},
  booktitle = {Proceedings of the 39th International Conference on Machine Learning},
  series = {Proceedings of Machine Learning Research},
  volume = {162},
  pages = {1298--1312},
  year = {2022},
  publisher = {PMLR}
}

@article{tang2026eegcapture,
  title = {What Do {EEG} Foundation Models Capture from Human Brain Signals?},
  author = {Tang, Ling and Chen, Qian and Mei, Jilin and Xu, Houshi and Zhang, Quanshi and Shao, Jing and Zou, Na and Hu, Xia and Liu, Dongrui},
  journal = {arXiv preprint arXiv:2605.11410},
  year = {2026}
}

@inproceedings{kornblith2019cka,
  title = {Similarity of Neural Network Representations Revisited},
  author = {Kornblith, Simon and Norouzi, Mohammad and Lee, Honglak and Hinton, Geoffrey},
  booktitle = {Proceedings of the 36th International Conference on Machine Learning},
  series = {Proceedings of Machine Learning Research},
  volume = {97},
  pages = {3519--3529},
  year = {2019},
  publisher = {PMLR}
}

\end{document}